**Influence of pressure on the magnetic behavior and the anomalous magnetoresistance in Tb$_5$Si$_3$**

Niharika Mohapatra, Sitikantha D Das, K. Mukherjee, Kartik K Iyer, and E.V. Sampathkumaran[*]
*Tata Institute of Fundamental Research, Homi Bhabha Road, Colaba, Mumbai 400005, India*

The compound, Tb$_5$Si$_3$, crystallizing in Mn$_5$Si$_3$-type hexagonal structure, was recently reported by us to exhibit a sudden and huge enhancement in electrical resistivity ($\rho$) at a critical magnetic field ($H_{cr}$) in the magnetically ordered state (< 70 K) tracking isothermal magnetization ($M$) behavior. We have investigated the influence of external pressure (≤15 kbar) and negative chemical pressure induced by Ge substitution for Si on $M$ and $\rho$ as a function of temperature (5-300 K) and magnetic field (<120 kOe), with the primary aim of understanding the field-induced anomalies. Focusing on isothermal $M$ and magnetoresistance (MR) at two temperatures, 5 and 20 K, we find that this $\rho$ anomaly persists under external as well as negative chemical pressures, however, with a change in the $H_{cr}$. The pressure-derivative of $H_{cr}$ is negative and this trend and the MR behavior at the $H_{cr}$ are comparable to that observed in some Laves phase itinerant magnetic systems. On the basis of this observation, we speculate that the magnetic fluctuations induced at this critical field could be responsible for the MR anomalies.

# I. INTRODUCTION

The process of metamagnetism [1], in which a low magnetic moment state of a magnetic material gets transformed to a high magnetic moment state with the application of an external magnetic field ($H_{ext}$), continues to attract attention in condensed matter physics. Some phenomena like 'giant magnetoresistance' and the 'giant magnetocaloric effect', which are of great current interest, are closely associated with this process. In particular, for the case of rare-earth ($R$) intermetallics, such a field-induced change in magnetism has been widely reported in the literature. In these local-moment systems, the electrical resistivity ($\rho$) at the metamagnetic transition usually decreases resulting in negative magnetoresistance, MR [defined as $\{\rho(H)-\rho(0)\}/\rho(0)$], due to a reduction in magnetic scattering by the application of magnetic field ($H$). The reader may see, for instance, Ref. 2 for the behavior of the MR in $Gd_5Si_2Ge_2$ and in DySb following the metamagnetic transition. In contrast to this general behavior, recently we observed [3] an enhanced positive MR at the field-induced magnetic transition in the case of a local-moment system, $Tb_5Si_3$. We considered it important to obtain more experimental results on this compound to understand the origin of such a MR anomaly.

At this juncture, it is to be noted that the itinerant systems [4] exhibiting the field-induced magnetic anomalies gained considerable theoretical and experimental interest. In particular, the above mentioned MR anomaly was in fact observed in the cubic Laves phase family, $RCo_2$. These binary compounds have been subjected to intense experimental investigations over a period of two decades [5-9]. In these systems, the existence of field-induced transitions and MR anomalies is known to originate from the existence of a peak in the Co 3d density of states in the vicinity of Fermi level. Yamada [7] theoretically predicted that the metamagnetic transition field ($H_t$) increases with increasing pressure ($P$) for such itinerant systems. For a case where $R$ carries a magnetic moment, below the ferromagnetic ordering temperature of the $R$ sublattice, the Co sublattice also orders due to a molecular field ($H_R$). For instance, when $R=Er$, one observes the magnetic transitions of Er and Co essentially at the same temperature and the Co moment couples antiparallel to Er moment. The magnitude of the Co moment is small compared to that on Er and the orientation of the Co moment is opposite to that of $H_{ext}$. For such antiferromagnetically-coupled itinerant systems, if $H_R$ is tuned by gradually replacing Er by Y, above a certain critical value of $H_{ext}$ (in the magnetically ordered state) satisfying the condition $H_R - H_{ext} \leq H_t$, the itinerant magnetism of the Co collapses and paramagnetic fluctuations are introduced in the Co sublattice. This phenomenon is called 'inverse metamagnetism'. Therefore, just above the critical $H_{ext}$, satisfying the above condition, $\rho$ should increase, instead of the decrease usually seen at the metamagnetic transition, thereby resulting in positive MR. The $H_{ext}$ required to bring about this phenomenon in such systems should actually decrease with increasing pressure within the theory of Yamada [7] and these ideas have been extensively discussed and demonstrated to explain the magnetic behavior in the solid solutions derived from $ErCo_2$ [9, 10]. However, to our knowledge, an analogy in local moment metallic antiferromagnetic systems had not been discussed in the past literature to infer the onset of such magnetic fluctuations and consequent influence on MR at a critical external field.

In view of the resemblance of the MR behaviors of $Er_{0.7}Y_{0.3}Co_2$ [Ref. 9] and $Tb_5Si_3$ [3], we considered it important to subject the latter for further studies, as stated in the first paragraph, to probe the influence of external pressure and negative chemical pressure (substitution of Ge for Si). This is the primary motivation of the present investigation. This compound has been known to crystallize in $Mn_5Si_3$-type hexagonal structure (see, figure 1, space group $P6_3/mcm$) [11-13]. In this crystal structure, there are two sites for Tb, viz., 4$d$ and 6$g$. This compound was reported to order magnetically near ($T_N$ ~) 70 K with a complex antiferromagnetic order (Semitelou et al, [12]). We have earlier reported [3] that, in the isothermal magnetization (*M*) data in the magnetically ordered state, the applied magnetic field required (which in this case is denoted by $H_{cr}$ rather than by $H_t$) to induce the transition decreases with increasing temperature (*T*) with the persistence of this transition and positive MR anomaly at temperatures very close to $T_N$ (for instance, at 60 K).

This article reports how the features near $T_N$ and the irreversibility of *M(H)* and MR(*H*) curves are influenced by external and negative chemical pressures. The central finding is that the pressure dependence of $H_{cr}$ in $Tb_5Si_3$ interestingly resembles the behavior seen in $Er_{0.7}Y_{0.3}Co_2$, despite the fact that the former is not an itinerant system. We demonstrate this conclusion by focusing our attention on the isothermal *M* and MR behavior at two temperatures, 5 and 20 K.

## II. EXPERIMENTAL DETAILS

Polycrystalline samples of the series, $Tb_5Si_{3-x}Ge_x$ (*x*= 0, 0.6, 1.5 and 2.0), were prepared by arc melting stoichiometric amounts of high purity (>99.9%) constituent elements in an atmosphere of high purity argon. The samples thus obtained were found to be single phase by x-ray diffraction (Cu $K_\alpha$) and the unit-cell volume (*V*) undergoes a monotonic increase with increasing Ge concentration (as expected) as shown in figure 2. The homogeneity of the specimens was ensured by scanning electron microscope and the compositions were ascertained by energy dispersive x-ray analysis. Magnetization as a function of temperature and magnetic field were performed with the help of a commercial SQUID magnetometer (Quantum Design) and a commercial vibrating sample magnetometer (Oxford Instruments). The ρ measurements in the presence of magnetic fields were performed by a commercial physical property measurements system (PPMS) (Quantum Design) and the electrical contacts of the leads with the specimens were made by a conducting silver paint. The *dc* magnetization and $\rho$ measurements under pressure (≤10 kbar and ≤15 kbar respectively) were carried out employing a commercial cell (easy-Lab Technologies Ltd, UK) with the help of the above-mentioned SQUID magnetometer and PPMS instruments. The measurements were done in a hydrostatic pressure medium of Daphne oil in the case of magnetization studies and a mixture of pentane and iso-pentane in the case of $\rho$ studies. All the measurements were done for the zero-field-cooled (from 250 K) conditions of the specimens.

## III. RESULTS
### A. Magnetization behavior of $Tb_5Si_3$ under pressure

The temperature dependence of magnetization measured in a field of 5 kOe for various values of external pressure for the zero-field-cooled condition of the specimens is shown in figure 3. In order to highlight the features, the plots of *M(T)* are restricted to temperatures below 150 K. For the sake of clarity, the curves obtained in the high pressure experiments are shifted in the vertical axis. As reported previously in Ref. 3, there is a peak in the vicinity of 70 K with the peak spreading over about 8 K. It appears that a modest application of an external pressure (10 kbar) shifts the peak to a marginally lower temperature, which is highlighted by drawing a vertical dashed line in figure 3 at 70 K as a guide to the eye. But the peak appears to split for 10 kbar, as a result of which a shoulder appears near 75 K. This signifies that there could be an additional magnetic transition in the close vicinity of 70 K for this pressure.

In order to understand how the reported field-induced transition below 70 K is influenced by pressure, we show the *M(H)* curves in figure 4 at 5 and 20 K. This transition persists for all pressures at these temperatures. As noted for the curve for the ambient pressure conditions (see also Ref. 3), the plots obtained in high pressure experiments are hysteretic. The most important point to be inferred from figure 4 is that the field where the transition occurs shifts gradually toward a lower field with increasing pressure. For instance, the $H_{cr}$, obtained from the peak position in the plot of $dM(H)/dH$ versus $H$ is 58 kOe at 5 K and this field shifts to about 52 kOe for $P$= 10 kbar. Corresponding $H_{cr}$ values at 20 K are 52 kOe and 42 kOe respectively. A vertical dashed line is drawn in figure 4 at 58 kOe (which is $H_{cr}$ for 5 K) to show how the curves and transition fields shift with $P$. In fact, a similar decrease of critical field is distinctly visible for 20 K while reducing the field towards zero (see figure 4); however, corresponding features for $T$= 5 K get gradually smeared out with increasing pressure.

### B. Magnetization behavior of $Tb_5Si_{3-x}Ge_x$

In order to explore how negative chemical pressure influences the features in the magnetization, we show the $M(T)$ and $M(H)$ behaviors in figure 5 and 6 respectively for the Ge-substituted solid solution. From the $M(T)$ curves shown in figure 5, it is apparent that the expansion of the lattice induced by Ge substitution for Si results in a gradual upward shift of the magnetic ordering temperature, attaining a value close to 80 K for $Tb_5Ge_3$ [14]. This volume dependence follows the trend reported in Ref. 5 for other local moment systems. The point of note is that $M(H)$ curves (see figure 6) for 5 and 20 K reveal the existence of a field-induced transition for all compositions with the virgin curves showing a tendency to shift upwards with increasing Ge substitution. This trend is transparent in the transition field in the reverse leg of the hysteresis loops also. For $Tb_5Ge_3$, however, this $M(H)$ feature [15] surprisingly appears broadened in polycrystalline form.

Another finding we have made is that the hysteresis loop at 5 K appears to contract with increasing pressure and this will be further addressed later in this article.

### C. Electrical and magnetoresistance behavior of $Tb_5Si_3$ under pressure

Figure 7 shows the temperature dependence of the electrical resistivity in zero-field and in 50 kOe for selected external pressures. We have plotted normalized resistivity, as the absolute values are not very reliable due to microcracks in the specimens. It is distinctly seen that an upturn (marked by an arrow) sets in below 75 K, which was absent in the corresponding curve shown in Ref. 3. This discrepancy could perhaps be an experimental artifact, for instance, due to stress induced by the freezing of the pentane/isopentane solution in this temperature vicinity and/or condensation of trapped nitrogen gas. With the application of a magnetic field (50 kOe), this effect becomes more prominent in the present investigation, setting in at a marginally lower temperature. Other features are broadly the same as those discussed in Ref. 3, including negative temperature coefficient of $\rho$ at temperatures far away from $T_N$ [16] and positive magnetoresistance (till about 60 K). With increasing pressure, this crossover temperature at which MR changes sign is not influenced significantly. A drop in the range 10-15 K distinctly develops in the in-field $\rho(T)$ curves and it is not clear whether this is related to the proximity of the tricritical point (20 K, Ref. 3).

With respect to the $H$-dependence of MR under pressure, an enhancement of positive MR in the vicinity of $H_{cr}$, followed by a peak at a slightly higher field, is apparent in the curves shown in figure 8, as in the case of ambient pressure data. The peak values of MR in the virgin curve are close to 60% at 5 and 20 K. This peak is followed by a gradual decrease as the field is increased well above $H_{cr}$. The hysteretic nature of the MR curves in figure 8 persists under pressure as well, tracking *M(H)* curves, though hysteresis is much weaker at 20 K. For 5 K, in the reverse leg of field variation, MR follows closely the virgin curve initially; however, it continues to rise while decreasing *H* below $H_{cr}$, attaining a peak at much lower fields with a significantly large value of MR (close to 160%). This is found to be the case for all pressures. For this temperature, the value of MR falls on the virgin curve when H→0 if *P* is less than or equal to 10 kbar. At 15 kbar (at 5 K), it appears that the curve in the reverse leg does not merge with the virgin curve as H→ 0, thus making the hysteresis loop open. We believe that this behavior at 15 kbar is due to the existence of a significant fraction of the highly resistive high-field phase even after the field is reduced to zero, thereby resulting in phase co-existence phenomenon [17].

A point that emerges clearly from figure 8 is that the curves shift towards lower fields with increasing pressure, resembling the behavior of *M(H)* plots, both at 5 and 20 K. In other words, the field at which the MR suddenly gets enhanced decreases with increasing *P*. Correspondingly, there is a monotonic decrease of the transition field in the reverse leg as well. A finding of interest is that, at 5 K, the $H_{cr}$ (in the forward leg) as inferred from the MR curves decreases by about 6 kOe as *P* is increased from 0 to 15 kbar, whereas, in the reverse leg, the corresponding change is much larger (> 20 kOe). In other words, the hysteresis loop at 5 K expands with increasing pressure, though it is not straightforward to draw this inference from the corresponding *M(H)* curves (figure 4). Such an expansion of hysteresis loop is not seen at 20 K. Clearly, this behavior occurs below the tricritical point.

### D. Electrical and magnetoresistance behavior of $Tb_5Si_{3-x}Ge_x$

The electrical transport behavior of Ge-substituted series is shown in figures 9 and 10. The gradual upturn in ρ(T) appearing in the paramagnetic state with decreasing temperature is essentially insensitive to Ge substitution. This increase of ρ as T → $T_N$ with decreasing temperature exhibits a tendency to get suppressed by the application of 50 kOe. Thus, the above upturn is almost absent for *x*= 0.6 -3.0 when ρ is measured in such a high-field. If one tracks the peak temperature in ρ(T) as well as the temperature at which there is a sharp fall (just below the peak) in the zero-field curves, the features confirm the findings from the magnetization data that $T_N$ undergoes a systematic increase with the expansion of the lattice with a value of about 85 K for $Tb_5Ge_3$. Thus, $T_N$ variation in this solid solution follows the trend observed in the external pressure experiments. It is obvious from figure 9 that the positive MR persists almost till $T_N$ for all compositions. One can also infer from figure 9 that the $T_N$, as inferred from the temperature at which ρ drops, gets marginally depressed for *H*= 50 kOe and this conclusion is most transparent for Ge-rich alloys.

In figure 10, we show MR as a function of *H* at 5 and 20 K for these alloys. As in the parent compound, following the field-induced transition, MR gets enhanced to a peak value and then exhibits a gradual decrease in its magnitude, interestingly for all compositions including the end member $Tb_5Ge_3$. It is to be noted that, for $Tb_5Ge_3$, unlike in the *M(H)* data presented above, the field-induced transition appears very prominently in the form of a significant change of slope, for instance, near 80 and 75 kOe at 5 and 20 K respectively. Possibly, the electrical resistivity is strongly anisotropic and, in polycrystals, electrical conduction is presumably dominated along the direction exhibiting these transitions. It is apparent from this figure that the field at which sharp upturn in the virgin curve occurs increases monotonically as one increases

Ge concentration, as in $M(H)$ curves discussed above. Needless to mention that the variation of MR with $H$ gets more hysteretic with decreasing temperature, which is obvious from a comparison of the curves in figure 10 for 5 and 20 K. The trend described above in the transition field with varying Ge composition in the virgin curves is seen while reducing the field as well. Like in $M(H)$ curves (figure 6), with increasing Ge concentration, the size of the hysteresis loop (figure 10) for 5 K appears to get contracted. This means that the transition field in the reverse leg undergoes a stronger increase with $x$ compared to that of $H_{cr}$ in the forward cycle. Clearly, negative chemical pressure extrapolates the behavior observed in external pressure experiments as far as the hysteresis loop behavior is concerned.

## IV. Discussion

In figure 11, we have compiled $H_{cr}$ obtained from the above measurements. For this purpose, $H_{cr}$ is defined as the peak field in the plots of $d(MR)/dH$ and $dM/dH$ versus $H$. Clearly, the variation of $H_{cr}$ in the solid solution is consistent with the trend seen in the external pressure experiments, compelling us to conclude that $H_{cr}$ decreases with decreasing volume. The positive volume coefficient of $H_{cr}$ interestingly follows the behavior seen in inverse metamagnetic itinerant systems [8, 9]. This naturally raises a question whether magnetic fluctuations as in 'inverse metamagnetism' can be invoked for the observed jump in positive MR. We make the following speculative points in support of this, though additional studies (for instance, neutron diffraction) are warranted to clarify further. In the parent Tb compound, there are two sites for Tb and the molecular field due to one site is presumably influencing magnetic order at the other site favoring an antiferromagnetic component between these two sites, though the net magnetic structure could be more complex [12]. As the magnetic field is increased in the magnetically ordered state (for the parent compound), $H_{ext}$ tries to disturb the two sub-lattice coupling, as a result of which there are possibly magnetic fluctuations induced (at least) at one of the sites, resulting in a higher resistive state at $H_{cr}$. In fact, we have also performed similar studies on Lu substituted series, $Tb_{5-x}Lu_xSi_3$, and we found (not shown here) that the sharp field-induced transition and the positive MR anomaly indeed vanish for 40 atomic percent of Lu (in the place of Tb), possibly due to a reduction in the molecular field and consequent weakening of intersite magnetic coupling. Additionally, we noted that, for $Tb_4LuSi_3$, at 2 K, the ρ for the 'supercooled' magnetic phase obtained by cycling through $H_{cr}$ exhibits quadratic field dependence in the range 0 - 50 kOe, which is typical of paramagnetic fluctuations. In fact, even for $Tb_5Si_3$, at 1.8 K for 10 kbar pressure [18], the transition in the reverse leg of field-variation could be suppressed and the super-cooled magnetic phase thus obtained exhibits a similar quadratic field-dependence in $MR(H)$; we argued that consistent interpretation of the observed trends in $M(H)$ and $MR(H)$ as $H \rightarrow 0$ for the supercooled magnetic phase is possible, only if we one assumes that the phase beyond $H_{cr}$ is characterized by magnetic fluctuations [18] and not by ferromagnetic alignment as in metamagnetism. The continuous decrease of MR with increase of $H$ well beyond the transition field in all our alloys is also a characteristic feature of suppression of magnetic fluctuations with the application of magnetic fields. During publication process of this paper, we came to know that, in another local moment antiferromagnetic system $PrSb_2$, a similar jump in positive MR and pressure dependence of $H_{cr}$ have been known [19]. Therefore such a MR anomaly must be more widespread under favorable circumstances in local moment antiferromagnets. It is possible that the underlying principle for the origin of such MR anomalies is the two (or more) sub-lattice (antiferro)magnetic structure, in which the magnetic fluctuations are induced at a critical external field. It is implicit in our speculation that it is not necessary to have high density of states (associated with itinerant systems discussed above) to induce the magnetoresistance anomalies under discussion. Within this scenario, it is an open question what decides the magnitude of $H_{cr}$.

Another observation we have made is that there is an expansion (a contraction) of the hysteresis loops with increasing pressure (negative chemical pressure). Such an expansion of the hysteresis loop as a function of temperature has been recognized across first-order magnetic transitions in some systems [17, 20-23] and it may have implications for magnetic phase co-existence phenomenon as discussed in references 22 and 23.

## V. CONCLUSIONS

We have investigated magnetization and transport anomalies in $Tb_5Si_3$ through external pressure and solid solution studies. The magnetic-field-induced anomalies do not vanish due to lattice compression or lattice expansion. Interesting irreversibility behavior of MR($H$) and $M(H)$ at 5 K as a function of external and chemical pressures pointing towards an expansion of hysteresis loop with decreasing volume have been observed. An important outcome of this work is that the pressure dependence of the critical field required to induce a magnetic transition with an enhancement in positive MR mimics that observed for an itinerant moment system exhibiting inverse metamagnetism. On the basis of this comparative study, we speculate that the magnetic fluctuations presumably induced by magnetic field on the antiferromagnetic structure could be a source of the MR anomalies under discussion. Hence, this work suggests the need to devise new theoretical approaches, particularly to derive $H_{cr}$, that are applicable for local moment metallic systems to understand such an anomalous field-induced transition and resultant transport anomalies.

*[Note added to cond-mat version:* At 1.8 K, a careful look at the *M(H)* data at for $Tb_5Si_3$ in our earlier paper (Ref. 3) reveals that there are additional weak steps beyond 58 kOe. We find that these steps get prominent with Ge substitution. We will address this issue in a forthcoming publication.]

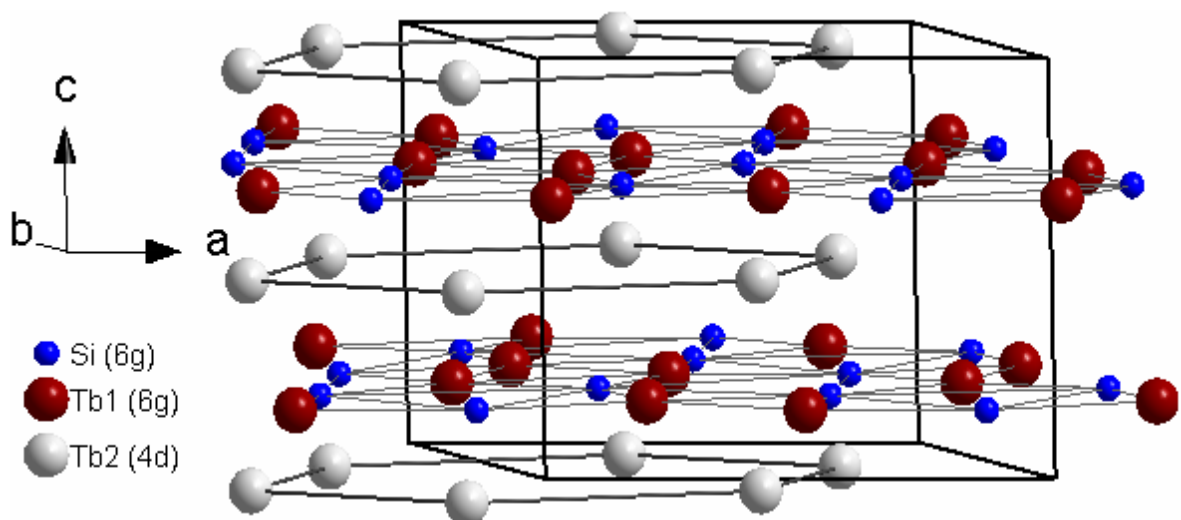

Figure 1:
(color online) Crystal structure of $Tb_5Si_3$.

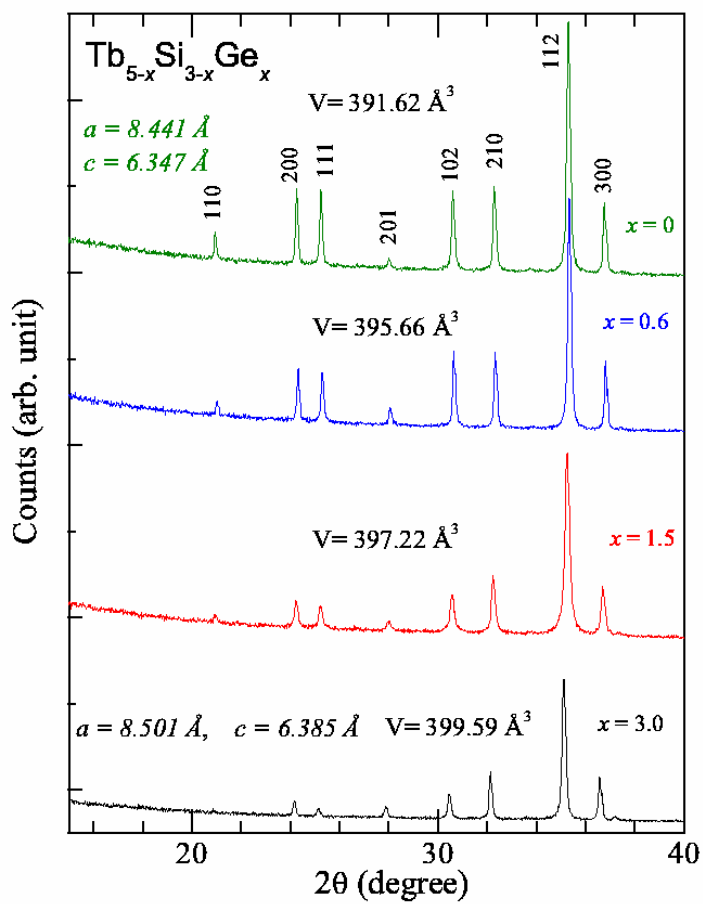

Figure 2:
(color online) X-ray diffraction patterns below $2\theta = 40°$ for the alloys, $Tb_5Si_{3-x}Ge_x$. The lattice constants, $a$ and $c$ ($\pm$ 0.004 Å), for end members and the unit-cell volume (V) for all compositions are also included. The curves are shifted along $y$-axis for the sake of clarity.

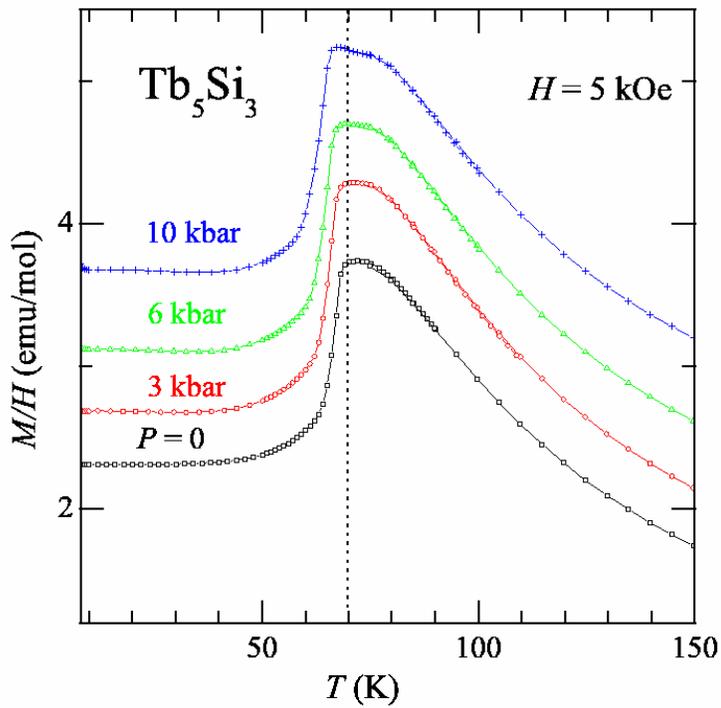

Figure 3:
(color online) Magnetization divided by magnetic field obtained in a field of 5 kOe under the influence of external pressure for $Tb_5Si_3$. A vertical dashed line at 70 K is drawn to serve as a guide to the eyes to show the shift of the peak with varying pressure. A line through the data points serve as guides to the eyes. The curves are shifted along *y*-axis for the sake of clarity.

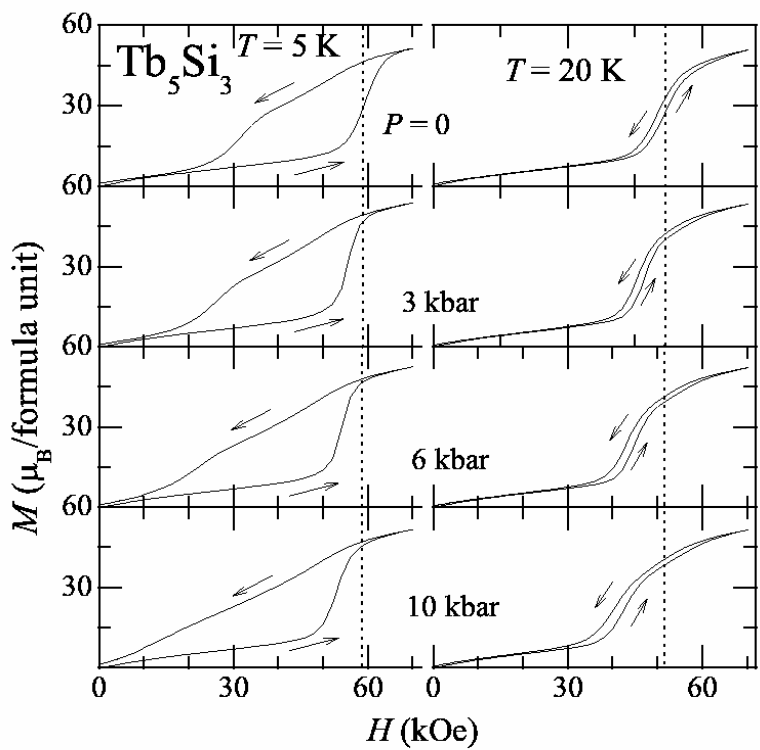

Figure 4:
Isothermal magnetization at 5 and 20 K under the influence of external pressure for $Tb_5Si_3$. Arrows are drawn to serves as guides to the eyes. A vertical dashed line is drawn (as described in the text) to show the shift of the field at which *M* jumps in the upward cycle of *H,* as one varies pressure.

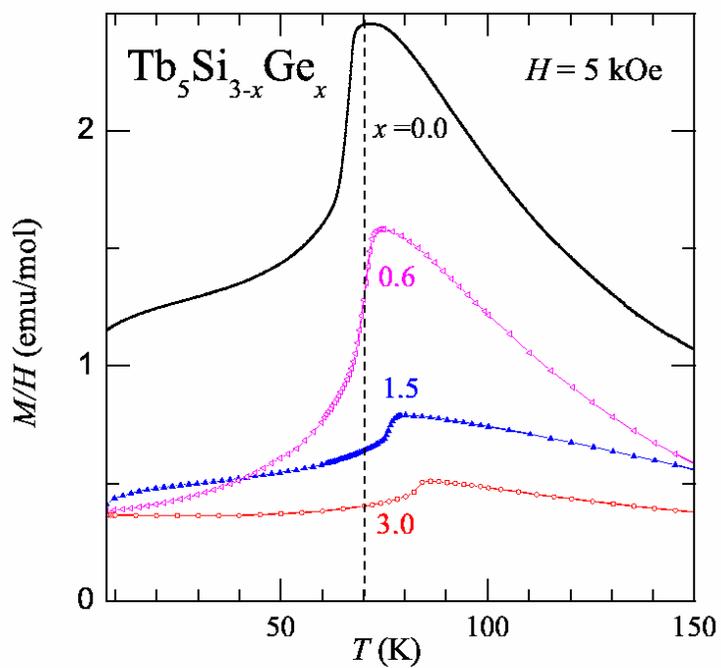

Figure 5:
(color online) Magnetization divided by magnetic field obtained in a field of 5 kOe for the alloys, $Tb_5Si_{3-x}Ge_x$. A vertical dashed line at 70 K is drawn to show the shift of the peak with varying composition. A line through the data points serve as guides to the eyes. The curves are shifted along $y$-axis for the sake of clarity.

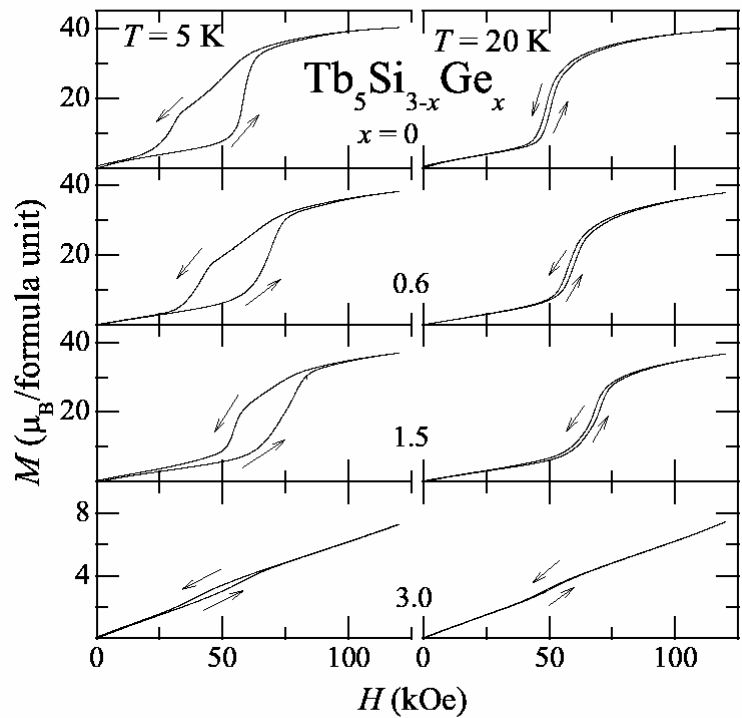

Figure 6:
Isothermal magnetization at 5 and 20 K for the alloys, $Tb_5Si_{3-x}Ge_x$. Arrows drawn are guides to the eyes.

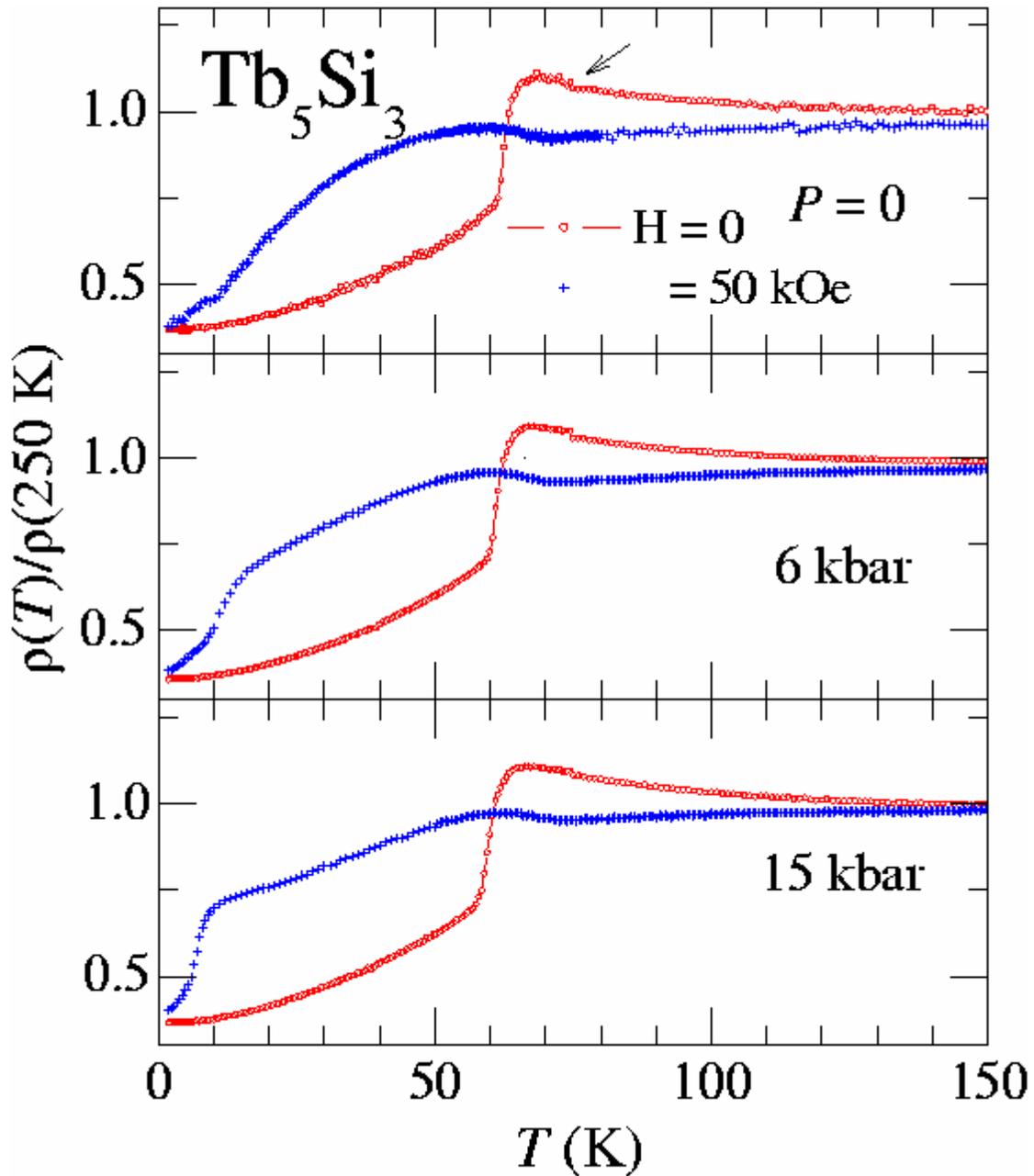

Figure 7:
(color online) Normalized electrical resistivity as a function of temperature in zero field and in the presence of 50 kOe for $Tb_5Si_3$ for selected pressures up to 15 kbar. Lines are drawn through the data points for zero field curves.

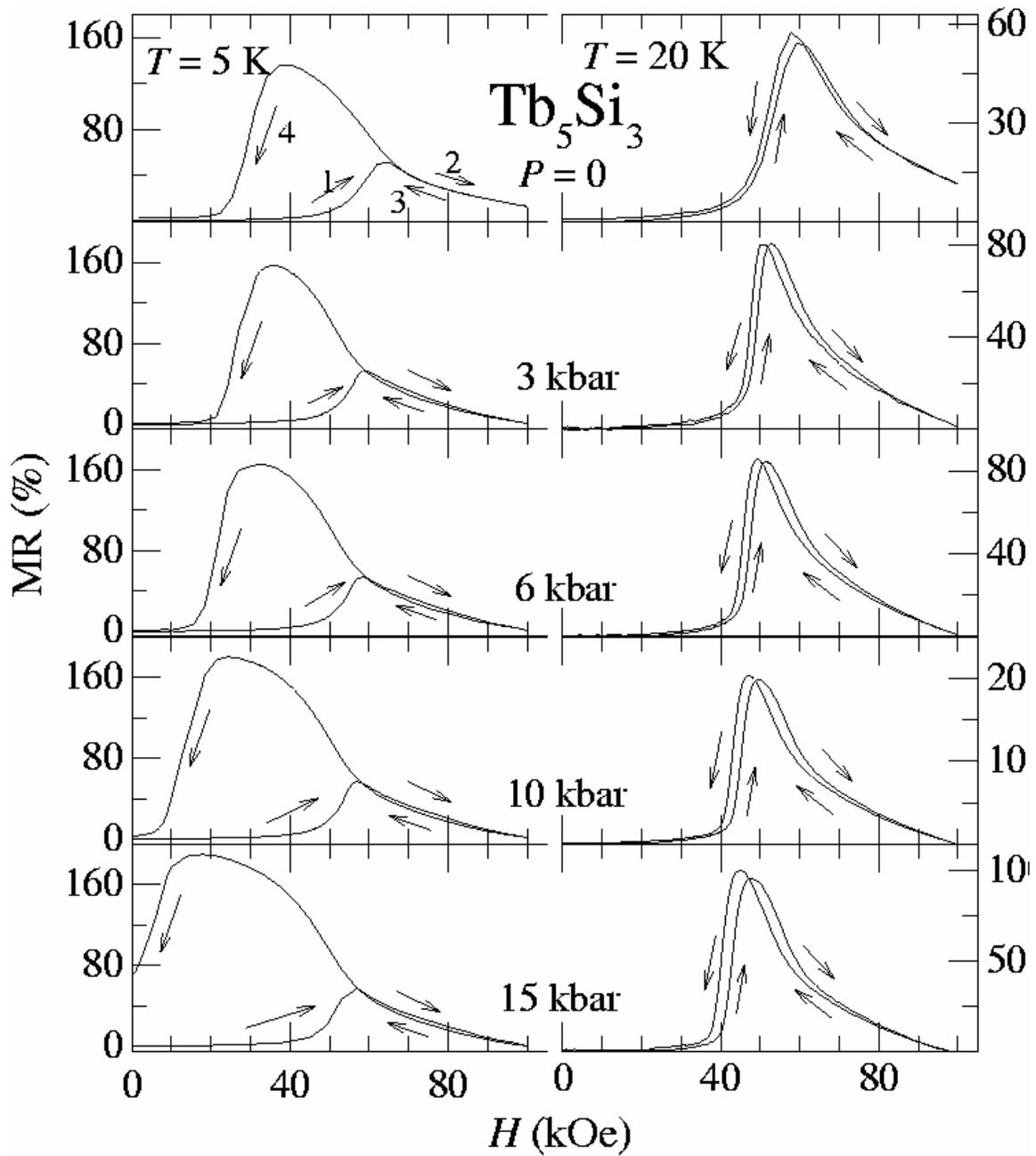

Figure 8:
Magnetoresistance at 5 and 20 K for $Tb_5Si_3$ for pressures up to 15 kbar. Arrows (and numericals for one curve) are placed to serve as guides to the eyes.

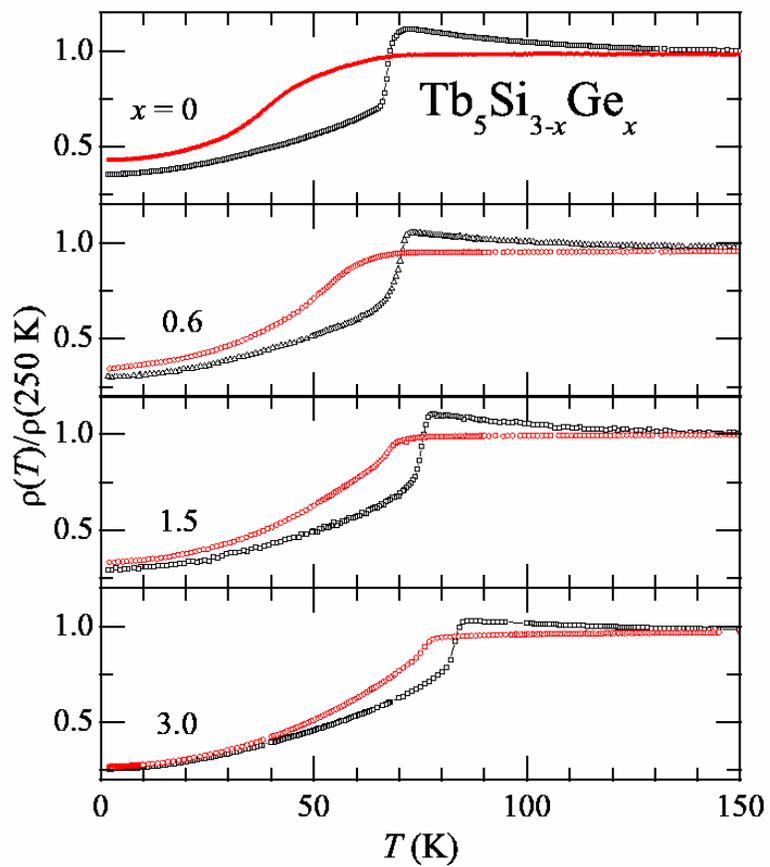

Figure 9:
(color online) Temperature dependence of normalized electrical resistivity for the alloys $Tb_5Si_{3-x}Ge_x$ in zero field and in the presence of 50 kOe. A line is drawn though the data points taken in zero-field.

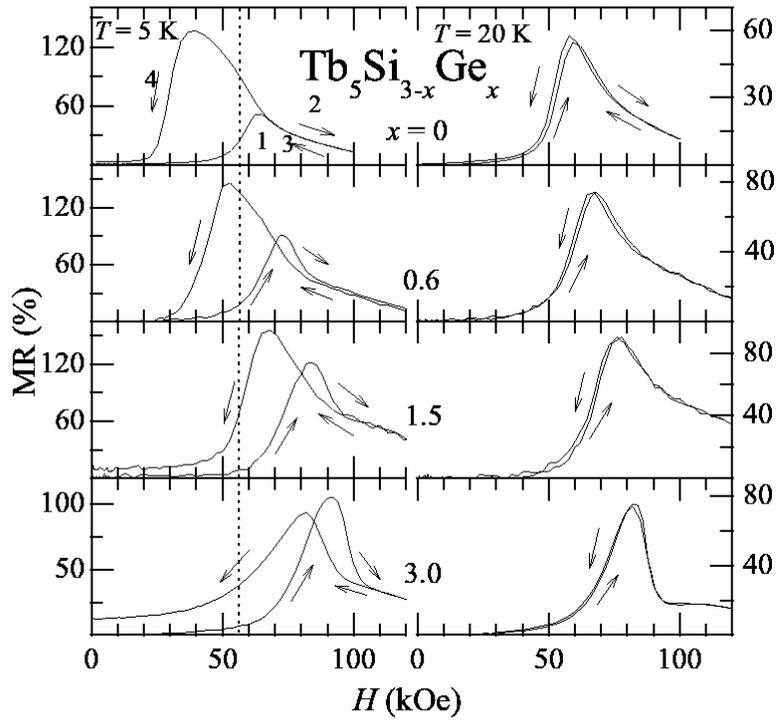

Figure 10:
Magnetoresistance as a function of magnetic field for the alloys $Tb_5Si_{3-x}Ge_x$ at 5 and 20 K. The arrows are marked to serve as guides to the eyes. A vertical dashed line for 5 K is drawn to show the shift of the field at which the MR jump occurs in the upward cycle of $H$ (as described in text), as one varies composition.

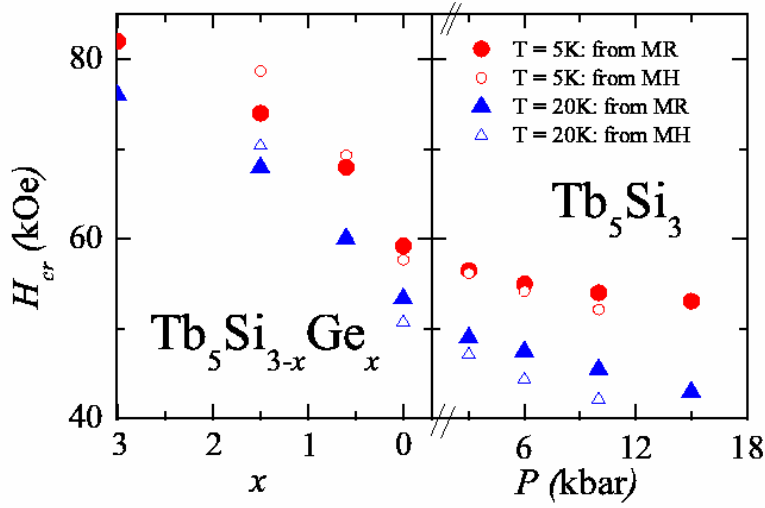

Figure 11: (color online) The plots of critical fields, $H_{cr}$, required to bring out a field-induced transition at 5 and 20 K, obtained from MR and $M$ data, as explained in the text.